\documentclass[aps,prl,twocolumn,superscriptaddress]{revtex4-1}
\usepackage{graphicx}

\begin{document}
\title{Antiferromagnetic order in epitaxial FeSe films on SrTiO3}

\author{Y. Zhou}
\email{equal contribution}
\affiliation{National Laboratory of Solid State Microstructures and Department of Physics, Nanjing University, Nanjing 210093, China}

\author{L. Miao}
\email{equal contribution}
\affiliation{Key Laboratory of Artificial Structures and Quantum
Control (Ministry of Education), School of Physics and Astronomy, Shanghai
Jiao Tong University, Shanghai 200240, China}

\author{P. Wang}
\affiliation{National Laboratory of Solid State Microstructures and Department of Physics, Nanjing University, Nanjing 210093, China}

\author{F. F. Zhu}
\author{W. X. Jiang}
\affiliation{Key Laboratory of Artificial Structures and Quantum
Control (Ministry of Education), School of Physics and Astronomy, Shanghai
Jiao Tong University, Shanghai 200240, China}

\author{S. W. Jiang}
\author{Y. Zhang}
\affiliation{National Laboratory of Solid State Microstructures and Department of Physics, Nanjing University, Nanjing 210093, China}

\author{B. Lei}
\author{X. H. Chen}
\affiliation{Hefei National Laboratory for Physical Sciences at Microscale and Department of Physics, University of Science and Technology of China and Key Laboratory of Strongly-coupled Quantum Matter Physics, Chinese Academy of Sciences, Hefei, Anhui 230026, China}
\affiliation{High Magnetic Field Laboratory, Chinese Academy of Sciences, Hefei, Anhui 230031, China}
\affiliation{Collaborative Innovation Center of Advanced Microstructures, Nanjing 210093, China}

\author{H. F. Ding}
\affiliation{National Laboratory of Solid State Microstructures and Department of Physics, Nanjing University, Nanjing 210093, China}
\affiliation{Collaborative Innovation Center of Advanced Microstructures, Nanjing 210093, China}

\author{H. Zheng}
\author{Jin-feng Jia}
\author{Dong Qian}
\email{dqian@sjtu.edu.cn}
\affiliation{Key Laboratory of Artificial Structures and Quantum
Control (Ministry of Education), School of Physics and Astronomy, Shanghai
Jiao Tong University, Shanghai 200240, China}
\affiliation{Collaborative Innovation Center of Advanced Microstructures, Nanjing 210093, China}

\author{D. Wu}
\email{dwu@nju.edu.cn}
\affiliation{National Laboratory of Solid State Microstructures and Department of Physics, Nanjing University, Nanjing 210093, China}
\affiliation{Collaborative Innovation Center of Advanced Microstructures, Nanjing 210093, China}

\date{\today}

\begin{abstract}

Single monolayer FeSe film grown on Nb-doped SrTiO$_3$(001) substrate shows the highest superconducting transition temperature (T$_C$ $\sim$ 100 K) among the iron-based superconductors (iron-pnictide), while T$_C$ of bulk FeSe is only $\sim$ 8 K. Antiferromagnetic spin fluctuations were believed to be crucial in iron-pnictides, which has inspired several proposals to understand the FeSe/SrTiO$_3$ system. Although bulk FeSe does not show the antiferromagnetic order, calculations suggest that the parent FeSe/SrTiO$_3$ films are AFM. Experimentally, due to lacking of direct probe, the magnetic state of FeSe/SrTiO$_3$ films remains mysterious. Here, we report the direct evidences of the antiferromagnetic order in the parent FeSe/SrTiO$_3$ films by the magnetic exchange bias effect measurements. The phase transition temperature is $\geq$ 140 K for single monolayer film. The AFM order disappears after electron doping.

\end{abstract}


\maketitle

The pairing mechanism of high-temperature superconductors including cuprates and iron-pnictides is one of the biggest challenges in modern physics. The antiferromagnetic (AFM) interaction has been long thought to correlate with the high-temperature superconductivity\cite{3,13} because the superconducting state usually appears after the AFM order is suppressed\cite{5,14}. The AFM spin fluctuations were proposed to play an important role in pairing in iron-pnictides\cite{3,4,5}. Among various iron-pnictides, FeSe has the simplest crystalline structure\cite{2}. T$_C$ of bulk FeSe is $\sim$ 8 K and can increase to $\sim$ 37 K under high pressure\cite{9}. Unlike other iron-pnictides, bulk FeSe crystals do not show AFM order\cite{9} unless applying certain pressure\cite{15,16,17}.

Surprisingly, the single monolayer (1-ML) FeSe film grown on Nb-doped SrTiO$_3$(001) ("STO" refer to Nb-doped SrTiO$_3$(001)) substrate after electron doping (through annealing process) shows a very large superconducting gap ($\sim$ 20 meV)\cite{18}. This gap survives up to $\sim$ 65 K\cite{12,19}. The diamagnetic signals below $\sim$ 65 K were also reported\cite{20}. Recently, the \textit{in situ} resistance measurements showed that T$_C$ of the 1-ML FeSe/STO film can be as high as 109 K\cite{1}. The mechanism of such high T$_C$ is still an open question. Several models were raised to understand the exotic properties of the 1-ML FeSe/STO film. Calculations showed that the electron-phonon coupling is significantly enhanced due to the interfacial effect in this system, but cannot explain such high T$_C$ if the paring is solely phonon-mediated\cite{21}. Another model calculation suggested that the electron-phonon coupling could strongly enhance T$_C$ by assuming that the pairing is mediated by the spin fluctuations\cite{6}. First-principle calculations showed that the FeSe/STO interface could enhance the AFM interaction, which helps maintain large spin fluctuations under heavy electron doping\cite{7}. Magnetic frustration induced by the combination of the electron doping and phonons is another possible mechanism for the superconductivity\cite{8}. Density functional theory (DFT) calculations suggested that the magnetic ground state of the 1-ML FeSe/STO film is AFM order\cite{7,10,11}. Recent work also claimed 1-ML FeSe/STO could be in AFM order to form topological superconductivity\cite{Liufeng}. Therefore, it's very interesting to study the magnetic ground state of the 1-ML FeSe/STO film before electron doping. Experimentally, the magnetic state of the FeSe/STO films is barely known. Previous angle-resolved photoemission spectroscopy (ARPES) measurements showed indirect and preliminary signatures of the spin density wave\cite{12}, but it is indistinguishable from the effect of the electronic nematicity\cite{22,23,24}. To determine the magnetic state, regular techniques such as neutron scattering, muon spin rotation and M\"{o}ssbauer effect have limited sensitivity for ultrathin films. In this work, we present the direct evidences that the magnetic ground state of the parent 1-ML FeSe/STO film is AFM by using the magnetic exchange bias effect (MEBE)\cite{25} measurements.

FeSe/STO films were grown as the previous reports\cite{16,1}. Films before post-annealing are called "as-grown" films. In order to become superconducting, as-grown films were post-annealed at $\sim$ 500$^o$C for 4 $\sim$ 8 hours \textit{in situ}. We call annealed films as "annealed" films. Before the films were transferred to another chamber to grow Fe$_{21}$Ni$_{79}$ layer, a 50-nm-thick Se protecting layer was grown. The polycrystalline Fe$_{21}$Ni$_{79}$ film was grown on the FeSe film at room temperature by e-beam evaporation after removing Se protecting layer by properly annealing. The Fe$_{21}$Ni$_{79}$ thickness was optimized for the MEBE measurements. Finally, a 10 nm Au film was deposited to prevent the oxidation. Magnetic properties were measured by a Quantum Design SQUID-VSM system. The magnetic field (\textbf{H}) was set to zero in an oscillation mode to reduce the residual field of the magnet before measurements. The residual field was further calibrated by a reference sample of Au/Fe$_{21}$Ni$_{79}$ (10 nm)/STO (see Supplementary Materials (SM)). Details on how to determine the coercivity with high accuracy and the uncertainty is described in SM.

\begin{figure}[]
\includegraphics[width=7cm]{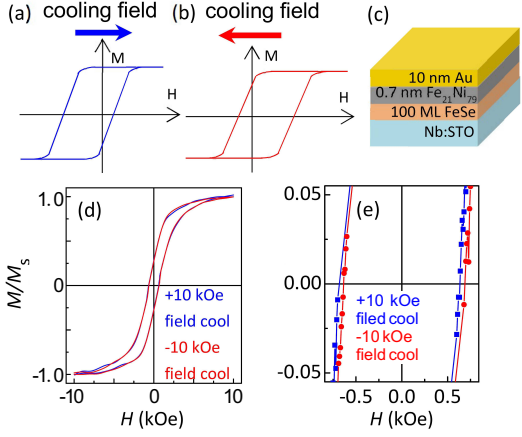}
\caption{Magnetic exchange bias effect in Fe$_{21}$Ni$_{79}$/FeSe/STO film. (a) and (b) The schematic magnetic hysteresis loops of the magnetic exchange bias effect after the positive and negative FC. (c)Layout of Au (10 nm)/Fe$_{21}$Ni$_{79}$ (0.7 nm)/FeSe (100 ML)/STO film.(d) Magnetic hysteresis loops of Au (10 nm)/Fe$_{21}$Ni$_{79}$ (0.7 nm)/FeSe (100 ML)/STO film measured at 5 K after FC (+/- 10 kOe) from room temperature to 5 K. (e) The corresponding zoom-in plots near the zero field. Biased loops were observed.}
\end{figure}

The MEBE is a magnetic effect which can be used for probing an AFM order in materials, particularly in thin films\cite{25,26}. The MEBE occurs in a ferromagnet/antiferromagnet heterostructure when it is cooled in an external \textbf{H} through the N\'{e}el temperature (T$_N$) of the AFM layer (with the Curie temperature of the ferromagnetic (FM) layer higher than T$_N$) or is grown in an external field. The MEBE relies on the interfacial magnetic exchange coupling between the AFM layer and the FM layer. The measurements are on the magnetization (\textbf{M}) of the FM layer. The distinct phenomenon of MEBE is that the center of the magnetic hysteresis loop (MHL) shifts away from the $\textbf{H}=0$, i.e., the absolute values of the coercive fields for increasing (H$_{C+}$) and decreasing (H$_{C-}$) fields are different. More importantly, the shifting direction reverses when the cooling field is reversed, as illustrated in Fig. 1(a) and 1(b).

First of all, we used thick FeSe/STO to show the capability of the MEBE measurements on FeSe system. We choosed the polycrystalline Fe$_{21}$Ni$_{79}$ as the FM layer. Figures 1(d) and 1(e) present the MHLs and the corresponding zoom-in plots of a Au (10 nm)/Fe$_{21}$Ni$_{79}$ (0.7 nm)/FeSe (100 ML)/STO sample measured at 5 K after field cooling (FC) from room temperature. The cooling field is either positive (blue curve) or negative (red curve) 10 kOe. The linear background originated from the diamagnetic signal of the STO substrate is subtracted from the raw data (see SM). The MHLs shift away from the zero field and the shifting direction is opposite to the direction of the cooling field. The shift of the MHLs and the reverse of the shifting direction upon reversing the cooling field direction indicate that the observed effect is MEBE. From Fig. 1(d), we obtain the magnitude of the shift -- exchange bias field ($H_{EB}) = \mid H_{C-}-H_{C+}\mid/2 \sim$ 28 Oe. Observed MEBE persists up to about 180 K in this sample.

\begin{figure}[]
\includegraphics[width=8.6cm]{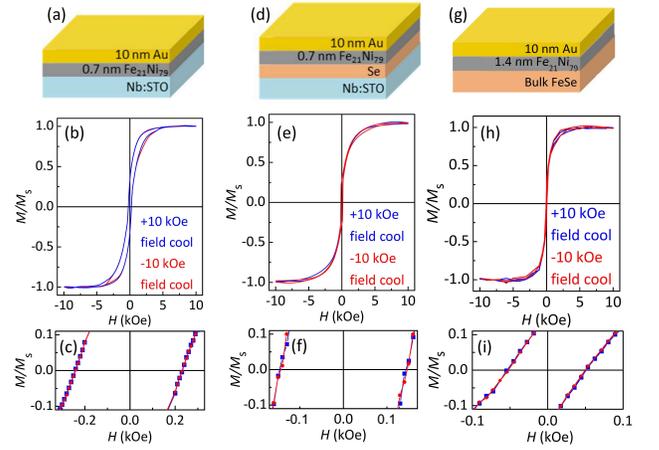}
\caption{Control experiments. (a) Layout of the Au (10 nm)/Fe$_{21}$Ni$_{79}$ (0.7 nm)/STO film. (b) Hysteresis loops of Au (10 nm)/Fe$_{21}$Ni$_{79}$ (0.7 nm)/STO film and (c) the corresponding zoom-in plots after FC. (d) Layout of the Au (10 nm)/Fe$_{21}$Ni$_{79}$ (0.7 nm)/Se (50 nm)/STO film. (e) Hysteresis loops of Au (10 nm)/Fe$_{21}$Ni$_{79}$ (0.7 nm)/Se (50 nm)/STO film and (f) the corresponding zoom-in plots after FC. (g) Layout of the Au (10 nm)/Fe$_{21}$Ni$_{79}$ (1.4 nm)/bulk-FeSe film. (h) Hysteresis loops of Au (10 nm)/Fe$_{21}$Ni$_{79}$ (1.4 nm)/bulk-FeSe film and (i) the corresponding zoom-in plots after FC. Loops did not shift in all control experiments.
}
\end{figure}

We carried out several control experiments to verify that the observed MEBE is the intrinsic property of the as-grown FeSe/STO films. i) First, we prepared a sample of Au (10 nm)/Fe$_{21}$Ni$_{79}$ (0.7 nm)/STO (Fig. 2(a)) and conduct the same measurements. Figures 2(b) and 2(c) show the MHLs and zoom-in plots. No shift was detected within our experimental uncertainty, which excludes any technical problems. ii) Before the deposition of the Fe$_{21}$Ni$_{79}$ film, FeSe/STO film was annealed to remove the Se protecting layer. Therefore, the Fe$_{21}$Ni$_{79}$ film might be selenited by the possible residual Se to form an AFlM layer at the interface and lead to MEBE. To exclude this possibility, we fabricated a control sample of Au (10 nm)/Fe$_{21}$Ni$_{79}$ (0.7 nm)/Se (50 nm)/STO (Fig. 2(d)). Figures 2(e) and 2(f) shows the MHLs and zoom-in plots. Clearly, loops do not shift, suggesting that the selenited Fe$_{21}$Ni$_{79}$ film is not AFM. iii) We also exclude that the AFM layer is caused by intermixing, alloying or proximity (polarization) effect between the Fe$_{21}$Ni$_{79}$ film and the FeSe film by using bulk FeSe as a reference. We prepared a sample of Au (10 nm)/Fe$_{21}$Ni$_{79}$ (1.4 nm)/bulk-FeSe (Fig. 2(g)). Here, a 1.4-nm-thick Fe$_{21}$Ni$_{79}$ film is used to obtain better signal-to-noise because the high quality cleaved surface area ($\sim$ 1$\times$1 mm$^2$) of bulk FeSe is much smaller than that of the STO substrate ($\sim$ 3$\times$3 mm$^2$). The measurement temperature (10 K) is slightly above T$_C$ of bulk FeSe to avoid the strong diamagnetic signal of the superconducting bulk FeSe. The cleaved bulk FeSe crystal has a (001) surface with Se-termination which is the same as the FeSe/STO film. Since both interfaces (Fe$_{21}$Ni$_{79}$/FeSe interface) are identical, one would expect the presence of MEBE in the Fe$_{21}$Ni$_{79}$/bulk-FeSe sample if the AFM layer is induced by the interfacial intermixing, alloying or polarization. However, MEBE is not observed in this control sample, shown in Figs. 2(h) and 2(i). With three control experiments above, we conclude that the observed MEBE most likely originates from the AFM layer in the 100-ML FeSe/STO film. In addition, for the completeness, we note that the MEBE can also occur between a heterostructure of a ferromagnet and a spin glass system\cite{25,26}. Although it is very unlikely to form the spin glass state in the FeSe/STO film, we carefully checked this possibility. The spin glass state in the FeSe/STO film is excluded by the thermal remnant magnetization measurements (see SM).

MEBE measurement can be used to determine the lower limit of T$_N$ of the AFM films. To show the temperature dependence more clearly, we used so called "inversion" method to plot the MHLs. In this method, both \textbf{M} and \textbf{H} of the original loop are multiplied by -1. The new loop is called the "inverted loop". After inversion, H$_{C-}$ of the original loop reflects from the negative \textbf{H} side to the positive \textbf{H} side, therefore we can directly show the difference between H$_{C+}$ and H$_{C-}$. Figures 3(a)-(f) show zoom-in original and inverted plots near the H$_{C\pm}$ of the Au (10 nm)/Fe$_{21}$Ni$_{79}$ (0.7 nm)/FeSe (100 ML)/STO sample measured at different temperature after FC. MEBE gradually becomes weak with increasing temperature. The temperature dependence of H$_{EB}$ is summarized in Fig. 3(g). We extracted the blocking temperature T$_B$, where H$_{EB}$ becomes zero, to be $\sim$ 180 K. The value of H$_{EB}$ depends on both AFM and FM layers, while T$_B$ mainly depends on the AFM layer\cite{25}. T$_B$ and T$_N$ are intimately correlated and in general T$_N\geq T_B$\cite{25}. Therefore, we obtained the lower limit of T$_N$ of $\sim$ 180 K for the 100-ML FeSe/STO film. The capability to detect the AFM order in iron-pnictides thin films by the MEBE is further demonstrated on the FeTe/STO film. FeTe has the similar crystal structure as FeSe and possesses a well-known AFM state\cite{27, 28}. On a Au (10 nm)/Fe$_{21}$Ni$_{79}$ (0.7 nm)/FeTe (50 ML)/STO sample, we observed MEBE. The determined T$_B$ of $\sim$ 75 K is comparable to the reported T$_N$ of thick FeTe films on MgO\cite{29} (see SM).

\begin{figure}[]
\includegraphics[width=8cm]{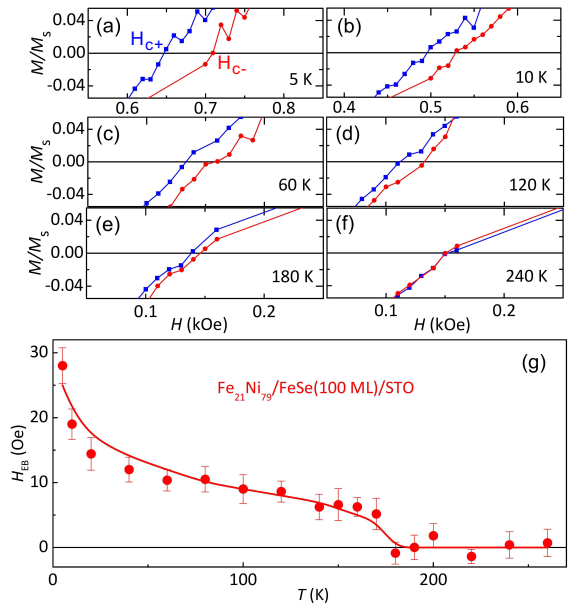}
\caption{Temperature dependence of MEBE. (a)-(f) The zoom-in curves of original (blue) and inverted loops (red) of Au (10 nm)/Fe$_{21}$Ni$_{79}$ (0.7 nm)/FeSe (100 ML)/STO film at different temperatures after FC. (g) H$_{EB}$ as a function of temperature. The red solid line is a guide for the eyes.
}
\end{figure}

\begin{figure*}[]
\includegraphics[width=12cm]{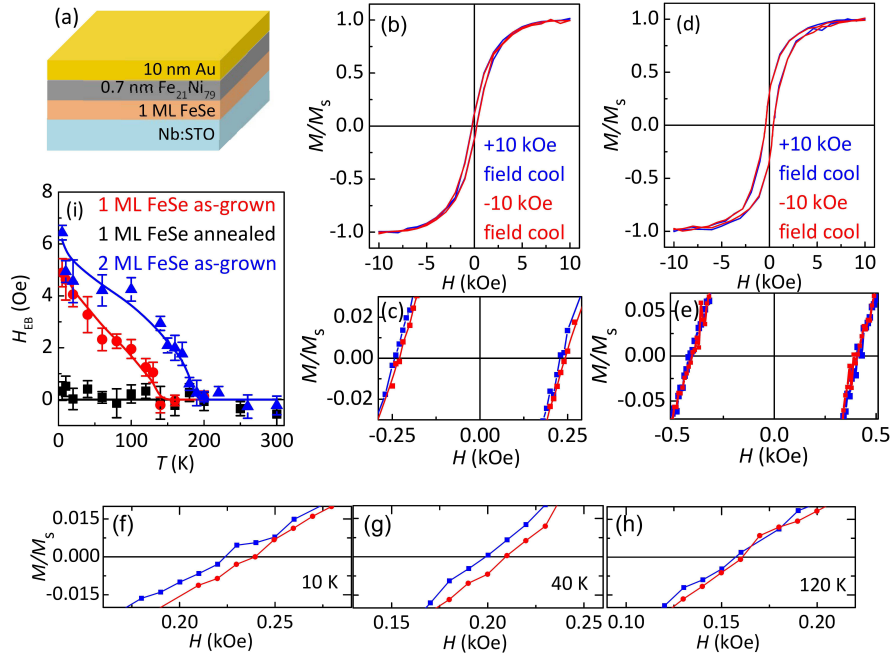}
\caption{MEBE in 1-ML and 2-ML FeSe/STO films. (a) Layout of Au (10 nm)/Fe$_{21}$Ni$_{79}$ (0.7 nm)/FeSe (1 ML)/STO film. (b) The magnetic hysteresis loops of Au (10 nm)/Fe$_{21}$Ni$_{79}$ (0.7 nm)/as-grown FeSe (1 ML)/STO film and (c) the corresponding zoom-in plots near zero field measured at 5 K after FC (+/- 10kOe) from room temperature to 5 K. (d) The magnetic hysteresis loops of Au (10 nm)/Fe$_{21}$Ni$_{79}$ (0.7 nm)/annealed FeSe (1 ML)/STO film and (e) the corresponding zoom-in plots near zero field measured at 5 K after FC (+/- 10kOe) from room temperature to 5 K. (f)-(h) The zoom-in curves of original (blue) and inverted loops (red) of Au (10 nm)/Fe$_{21}$Ni$_{79}$ (0.7 nm)/as-grown FeSe (1 ML)/STO film at three temperatures after FC. (g) H$_{EB}$ of the Au (10 nm)/Fe$_{21}$Ni$_{79}$ (0.7 nm)/as-grown or annealed FeSe (1 or 2 ML)/STO film as a function of temperature.
}
\end{figure*}

The chief motivation for this study is to check the magnetic state of the 1-ML FeSe/STO film. After the demonstration of the capability of the MEBE study on FeSe/STO as well as FeTe/STO films, we studied the 1-ML FeSe film. The as-grown (parent) 1-ML FeSe/STO film is non-superconducting. It becomes superconducting by doping electrons through the annealing process\cite{1,12,18,19,20}. We prepared two types of samples: Au (10 nm)/Fe$_{21}$Ni$_{79}$ (0.7 nm)/as-grown FeSe (1 ML)/STO (sample \#1) and Au (10 nm)/Fe$_{21}$Ni$_{79}$ (0.7 nm)/annealed FeSe (1 ML)/STO (sample \#2). Our sample \#1 is in the "N-phase" and sample \#2 is in the "S-phase" reported by Zhou's group\cite{19}. Superconducting gap was observed on annealed FeSe/STO films by ARPES (see SM). MEBE is clearly observed in sample \#1 at 5 K after FC, shown in Fig. 4(c). The shift of the MHL is relatively small ($\sim$ 5 Oe), but it is still well above the error bar ($\sim$ 0.5 Oe). In contrast, MEBE was not detected within our experimental uncertainty in sample \#2, shown in Fig. 4(e). The AFM order exists in the as-grown non-superconducting 1-ML FeSe/STO films and disappears on annealed films (heavy electron doping). Because we do not know whether the superconducting state in annealed 1-ML FeSe/STO film is maintained or not after interfacing with Fe$_{21}$Ni$_{79}$ film, we can not say that AFM order in as-grown sample is destroyed by superconductivity. We'd like to suggest that the heavy electron doping by annealing process destroys the AFM order. Figures 4(f)-(g) show zoom-in original and inverted plots near the H$_{C\pm}$ of sample \#1 measured at different temperature. Shown in Fig. 4(i), T$_B$ is $\sim$ 140 K, meaning T$_N\geq$ 140 K for the 1-ML as-grown FeSe/STO film. More data on different samples for the 1-ML films can be found in SM. T$_N$ of the as-grown 1-ML FeSe/STO film is much higher than the reported highest T$_N$ ($\sim$ 55 K) of the bulk FeSe under high pressure\cite{16}. Such high T$_N$ implies that it is possible to have Tc $\sim$ 109 K if spin-spin interaction plays crucial roles in electron's pairing together with the help of phonons suggested by Ref. 17.

MEBE dependents on the competition between the interfacial energy J$_{int}$ at the ferromagnet/antiferromagnet interface and the anisotropy energy K$_{AFM}$t$_{AFM}$ of the AFM layer, where K$_{AFM}$ and t$_{AFM}$ are the anisotropy constant and the thickness of the AFM layer, respectively. The condition K$_{AFM}$t$_{AFM}\geq$ J$_{int}$ or t$_{AFM}\geq$J$_{int}$/K$_{AFM}$ is required for the observation of the MEBE\cite{26,30}, meaning that a critical AFM thickness is needed for the MEBE\cite{26,30}. In Fe$_{21}$Ni$_{79}$/FeSe/STO system, J$_{int}$ is relatively weak because the interfacial coupling occurs indirectly between the Fe/Ni atoms of Fe$_{21}$Ni$_{79}$ and the Fe atoms of FeSe through the Se atoms, which means that the critical AFM thickness in Fe$_{21}$Ni$_{79}$/FeSe/STO system would be very thin. That is why we can observe the MEBE even in 1-ML-thick FeSe/STO film.

Furthermore, we carried out the measurements on the as-grown 2-ML FeSe/STO films. The Au (10 nm)/Fe$_{21}$Ni$_{79}$ (0.7 nm)/FeSe (2 ML)/STO sample exhibits the MEBE at low temperature after FC, indicating that the 2-ML FeSe/STO film also has the AFM order. T$_B$ is determined to be $\sim$ 180 K (Fig. 4(g)), which is larger than that of 1-ML film as expected due to the increased thickness of AFM layer\cite{25, 26}. Interestingly, T$_B$ of 2-ML FeSe/STO films is already similar to that of the 100-ML FeSe/STO film, which implies that the inter-layer magnetic interaction is much weaker than the intra-layer magnetic interaction compatible with the layered structure of FeSe.

Finally, we try to get some insight into why AFM order can exist in FeSe/STO films. Although the origin why bulk FeSe has no magnetic order is under debate\cite{Lee, Glasbrenner, Yu, Chubukov, Lu}, strong AFM spin fluctuations were observed in neutron scattering experiments\cite{Zhao1, Zhao2}. DFT calculation suggested that tensile stress could further enhance the AFM interaction\cite{7}. Is stress the reason why we observed AFM order? For thick films, the answer is very unlikely. 100-ML FeSe/STO films have very similar lattice constant as the bulk FeSe, but we still observed MEBE. Why thick films are different from bulk FeSe crystals? In fact, thick FeSe films have very different microscopic properties from bulk crystals. First, there are numbers of Fe vacancies in FeSe/STO films\cite{Xue}. Second, the strength of nematicity in FeSe/STO is much larger than that in bulk FeSe\cite{Xue, Zhang}. A very recent STM study on FeSe/STO films observed a stripe-type charge ordering that does not exist in bulk FeSe and the charge ordering is pinned in the vicinity of Fe vacancies as well as domain walls of nematicity\cite{Xue}. The pinned charge order is quantitatively comparable to a magnetic channel predicted by theory\cite{Ku}. In another word, impurities (or defects) could help to pin the magnetic fluctuations and form relatively long range AFM order. The existence of AFM order could be the reason why superconductivity does not recover in thick FeSe/STO films. On the other side, it is much more complicated in 1-ML films. Strong tensile stress and impurities (or defects) coexist\cite{18,1,TEM}. Tensile stress enhances the interaction, while impurities or defects help to pin AFM order, so we can not rule out any of them for 1-ML films. Annealing process can inject electrons into the first ML of FeSe. Interaction between local magnetic moments through mobile electrons would prefer FM state if Hund coupling dominates. Therefore, the competition between AFM and FM interactions could destroy the AFM order and eventually form superconductivity in 1-ML films. There might be other possibilities that can kill the AFM order during annealing process and more theoretical inputs are needed to fully understand the magnetic property of as-grown FeSe/STO films in the future.

In summary, we observed AFM order in parent 1-ML FeSe/STO films by MEBE measurements before electron doping. The low limit of the N\'{e}el temperature is about 140 K. The strong AFM interaction in the 1-ML FeSe/STO film could help the superconductivity if the AFM coupling plays an indispensable role in realizing superconducting state in 1-ML FeSe/STO film. Our findings provided very important information for the comprehensive understanding of the novel properties in FeSe/STO films.

We acknowledge Chunlei Gao, Wei Ku and Weijiong Chen for the experimental help and discussions. The work was supported by the Ministry of Science and Technology of China and NSFC. D.Q. acknowledges support from the Changjiang Scholars Program.

\end{document}